\documentstyle[psfig]{mn}
\newif\ifAMStwofonts
\ifoldfss
  \ifCUPmtlplainloaded \else
    \NewTextAlphabet{textbfit} {cmbxti10} {}
    \NewTextAlphabet{textbfss} {cmssbx10} {}
    \NewMathAlphabet{mathbfit} {cmbxti10} {} % for math mode
    \NewMathAlphabet{mathbfss} {cmssbx10} {} %  "   "    "
  \fi
  \ifAMStwofonts
    \ifCUPmtlplainloaded \else
      \NewSymbolFont{upmath} {eurm10}
      \NewSymbolFont{AMSa} {msam10}
      \NewMathSymbol{\upi}     {0}{upmath}{19}
      \NewMathSymbol{\umu}     {0}{upmath}{16}
      \NewMathSymbol{\upartial}{0}{upmath}{40}
      \NewMathSymbol{\leqslant}{3}{AMSa}{36}
      \NewMathSymbol{\geqslant}{3}{AMSa}{3E}

    \fi
  \fi
\fi % End of OFSS

\ifnfssone
  \newmathalphabet{\mathit}
  \addtoversion{normal}{\mathit}{cmr}{m}{it}
  \addtoversion{bold}{\mathit}{cmr}{bx}{it}
  \newmathalphabet{\mathbfit} % math mode version of \textbfit{..}
  \addtoversion{normal}{\mathbfit}{cmr}{bx}{it}
  \addtoversion{bold}{\mathbfit}{cmr}{bx}{it}
  \newmathalphabet{\mathbfss} % math mode version of \textbfss{..}
  \addtoversion{normal}{\mathbfss}{cmss}{bx}{n}
  \addtoversion{bold}{\mathbfss}{cmss}{bx}{n}
  \ifAMStwofonts
    \ifCUPmtlplainloaded \else
      \UseAMStwoboldmath
      \makeatletter
      \new@mathgroup\upmath@group
      \define@mathgroup\mv@normal\upmath@group{eur}{m}{n}
      \define@mathgroup\mv@bold\upmath@group{eur}{b}{n}
      \edef\UPM{\hexnumber\upmath@group}
      \new@mathgroup\amsa@group
      \define@mathgroup\mv@normal\amsa@group{msa}{m}{n}
      \define@mathgroup\mv@bold\amsa@group{msa}{m}{n}
      \edef\AMSa{\hexnumber\amsa@group}
      \makeatother
      \mathchardef\upi="0\UPM19
      \mathchardef\umu="0\UPM16
      \mathchardef\upartial="0\UPM40
      \mathchardef\leqslant="3\AMSa36
      \mathchardef\geqslant="3\AMSa3E
    \fi
  \fi
\fi % End of NFSS release 1												

\ifnfsstwo
  \DeclareMathAlphabet{\mathbfit}{OT1}{cmr}{bx}{it}
  \SetMathAlphabet\mathbfit{bold}{OT1}{cmr}{bx}{it}
  \DeclareMathAlphabet{\mathbfss}{OT1}{cmss}{bx}{n}
  \SetMathAlphabet\mathbfss{bold}{OT1}{cmss}{bx}{n}
  \ifAMStwofonts
    \ifCUPmtlplainloaded \else
      \DeclareSymbolFont{UPM}{U}{eur}{m}{n}
      \SetSymbolFont{UPM}{bold}{U}{eur}{b}{n}
      \DeclareSymbolFont{AMSa}{U}{msa}{m}{n}
      \DeclareMathSymbol{\upi}{0}{UPM}{"19}
      \DeclareMathSymbol{\umu}{0}{UPM}{"16}
      \DeclareMathSymbol{\upartial}{0}{UPM}{"40}
      \DeclareMathSymbol{\leqslant}{3}{AMSa}{"36}
      \DeclareMathSymbol{\geqslant}{3}{AMSa}{"3E}
    \fi
  \fi
\fi % End of NFSS release 2

\ifCUPmtlplainloaded \else
  \ifAMStwofonts \else % If no AMS fonts
    \def\upi{\pi}
    \def\umu{\mu}
    \def\upartial{\partial}
  \fi
\fi

\title{Spiral shocks in the accretion disc of IP Peg during outburst maximum}
\author[E.~T.~Harlaftis et al.]
       {E.~T.~Harlaftis$^1$\thanks{Previous address:
       School of Physics and Astronomy, University of St. Andrews, St 
       Andrews, KY16 9SS, UK}
	D. ~Steeghs$^2$,
        K.~Horne$^2$,
	E.~Mart\'{\i}n$^3$,
	and A.~Magazz\'{u}$^{4,5}$\\
$^1$Astronomical Institute, National Observatory of Athens, P.O. Box
20048, Athens 118 10, Greece\\
$^2$School of Physics and Astronomy, University of St. Andrews, KY16 9SS, 
Scotland, UK\\
$^3$Instituto de Astrof\'{\i}sica de Canarias, E-38200 La Laguna, Tenerife, 
Spain\\
$^4$Osservatorio Astrofisico di Catania, Citt\'{a} Universitaria,
I-95125, Catania, Italy\\
$^5$Centro Galileo Galilei, Apartado 565, E-38700 Santa Cruz de La Palma,
Spain}

\date{Accepted 21 January 1999; Received 23 August 1998; in original form 1998}
\pagerange{\pageref{firstpage}--\pageref{lastpage}}
\pubyear{1999}

\begin{document}

\maketitle

\label{firstpage}

\begin{abstract} 

In response to  our recent discovery of spiral  arms  in the accretion
disc of IP Peg during rise to outburst, we have obtained time-resolved
spectrophotometry of IP Peg  during outburst maximum.  In  particular,
indirect  imaging of  He{\small~II}   4686, using Doppler  tomography,
shows a  two-arm  spiral pattern on   the disc image,  which  confirms
repeatability over   different outbursts.  The  jump  in He{\small~II}
intensity (a factor of more than  two) and in velocity ($\sim$200--300
km s$^{-1}$) clarifies the shock nature  of the spiral structure.  The
He{\small~II}  shocks show an azimuthal  extent of $\sim$90 degrees, a
shallow   power-law emissivity $\sim~V^{-1}$,   an  upper limit of  30
degrees  in opening angle,  and a flux  contribution of 15 per cent of
the total disc  emission.  We  discuss the results  in view  of recent
simulations of accretion discs which  show that  spiral shocks can  be
raised in the accretion disc by the secondary star.

\end{abstract}

\begin{keywords}
cataclysmic variables, accretion disc, IP Pegasi
\end{keywords}

\section{Introduction}

\begin{figure}
%\vspace{140pt}
\psfig{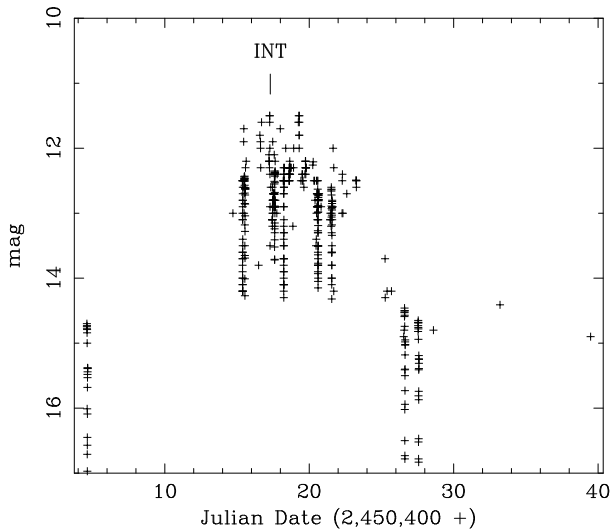}
\caption{The optical light curve, as given by AAVSO,
during the November 1996 outburst of IP Peg. 
Eclipses are visible as sharp drops in magnitude.
The time of the INT observations is marked.}
\end{figure}

High-energy   phenomena such  as jets  emanating   from AGNs and X-ray
radiation emitted   from binaries are   related  to  accretion  discs.
Accretion discs are the most efficient machines  for the extraction of
energy and angular   momentum.  The physical  conditions  that lead to
specific dynamical formations in  discs, such as  spiral arms, are not
contained  as  yet due to lack   of observational constraints.  Spiral
structure  in galaxies is attributed to  a quasi-stationery wave which
triggers star formation  as it propagates  through the disc (Bertin et
al.  1989) or   a tidal pattern   due to interaction  with a satellite
galaxy (e.g.  M51;  Toomre and Toomre 1972).  Spiral  waves in  a disc
have also been invoked to explain the proximity of giant extra-planets
to their  orbiting stars (e.g  51 Peg;  Lin  et al.  1996).  Moreover,
spiral shocks have  been found in  simulations of protosolar discs and
have been  proposed as a  means of converting the gaseous protostellar
disc into   orbiting planetesimals (Boss    1997).  Accretion discs in
cataclysmic variables (a white dwarf accretes matter  from a donor K-M
star) evolve on short dynamical timescales (2-10 hours) allowing us to
investigate  in detail  their development.  In  cataclysmic variables,
the   outburst origin in  dwarf novae  (disc  or  donor  star) and the
mechanism  for  the angular  momentum transport  of the  disc material
(`viscosity') are still   unresolved issues though  fundamental in our
understanding of   accretion   physics (Verbunt   1986).   The   exact
structure of the disc, an $\alpha$-disc  (Shakura and Synyaev 1973) or
a  spiral-shock disc (Sawada, Matsuda  \& Hachisu  1986), is linked to
the above  issues.  Simulations of such  accretion discs have revealed
double spiral shocks   (Sawada,  Matsuda \& Hachisu   1986;  Savonjie,
Papaloizou and   Lin 1994;  Stehle   1998). Recently,   we  indirectly
detected spiral  waves in the   accretion disc of the eclipsing  dwarf
nova IP  Peg during the  rise to  outburst (Steeghs,  Harlaftis, Horne
1997).  Here,  we  report on subsequent observations  of He{\small~II}
4686, during  maximum of the November  1996 outburst, which were aimed
to probe the ionization structure of the spiral arms. 

\section{Observations}

\begin{figure*} 
\psfig{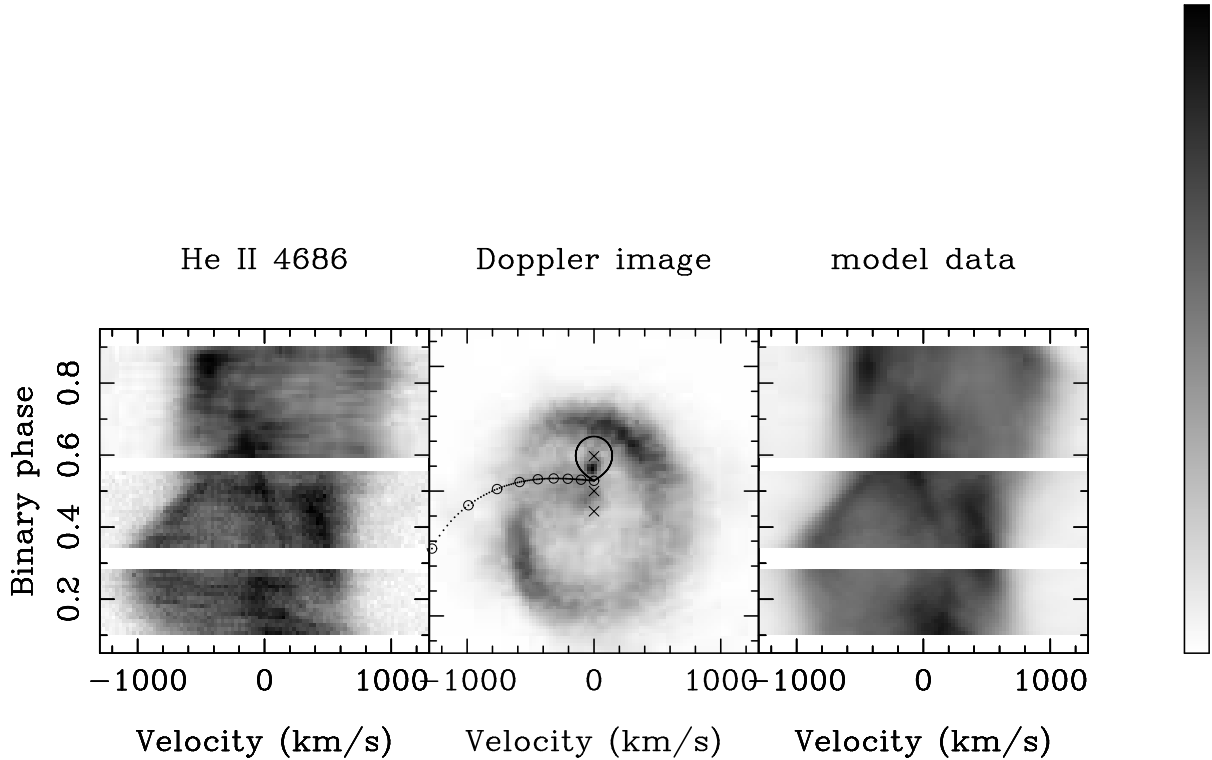}
\caption[]{Doppler tomography of He{\small~II} 4686.
From left to right, the panels display the 
observed data, the Doppler image, and the projection of the Doppler image 
into a trailed spectrogram.
The image shows non-axisymmetric components arising from the
red dwarf, the disc (tidal shocks) and an unresolved location with almost 
zero velocity. The red dwarf size and ballistic trajectory are also drawn
(for $K_{\rm c}=280$ and $K_{w}=162$ km s$^{-1}$).
Steps in units of $R_{L_{1}}$ are marked along the ballistic velocities.
The greyscale bar corresponds to a linear intensity scale of 0-57 mJy.}
\end{figure*}

IP Peg, a double-eclipsing dwarf  nova, shows semi-periodic  outbursts
every   $\sim$3 months.  During  the  third  day  of the November 1996
outburst (Fig. 1), we observed  a complete binary  cycle of 3.8 hours.
We  obtained 81 spectra, under  $\approx~1.6$ arcsecond seeing, with a
TEK CCD and the 235 camera of the Intermediate Dispersion Spectrograph
on the 2.5m Isaac Newton Telescope at La  Palma.  The wavelength range
covered is 4354--4747 \AA \ at a velocity dispersion of 27 km s$^{-1}$
per pixel.  The spectra were extracted using optimal extraction (Horne
1986) after  debiasing,     flat-fielding  and sky-subtraction.    The
wavelength calibration was performed using 17 CuAr arc lines and a 3rd
order polynomial fit and is accurate to  0.04\AA.  \ A comparison star
had been included in  the slit which  was subsequently used to correct
the object spectra for atmospheric and slit losses.  The absolute flux
scale was defined by observing  the flux standard Feige~15 (Oke 1990).
Typical signal to noise ratio is 21 for the IP Peg spectra (and 46 for
the  comparison  spectra).   Flux uncertainties were  estimated  using
Poisson statistics and  a noise model  for the CCD  chip  (gain of 1.1
electrons per ADU  and readout noise of 5.7  electrons per pixel).  We
adopt  the  binary ephemeris from  Wolf  et al. (1993)  $ T_{o}(HJD) =
2445615.4156(4) +   0.15820616(4)~E$,  where $T_{o}$ is  the  inferior
conjunction    of      the     white   dwarf.      The   He{\small~II}
$\lambda4542/\lambda4686$ flux ratio of  0.022-0.036,  as well as  the
He{\small~I}  flux   ratio  $\lambda6678/\lambda4471$  flux  ratio  of
0.71-0.87  (see also   Marsh and   Horne 1990),   are consistent  with
recombination case B  (for T = 5,000-20,000  K and electron density of
$10^{4}$ cm$^{-3}$; Osterbrock 1989).  Given  that the He{\small~I} is
blended with the Mg{\small~II}  line, the above  ratio is only a lower
limit, thus favouring the temperatures nearer to 20,000 K.

\section{Resolving Spiral waves with Doppler tomography}

\begin{figure*}
%\vspace{15cm}
%\psdraft
\psfig{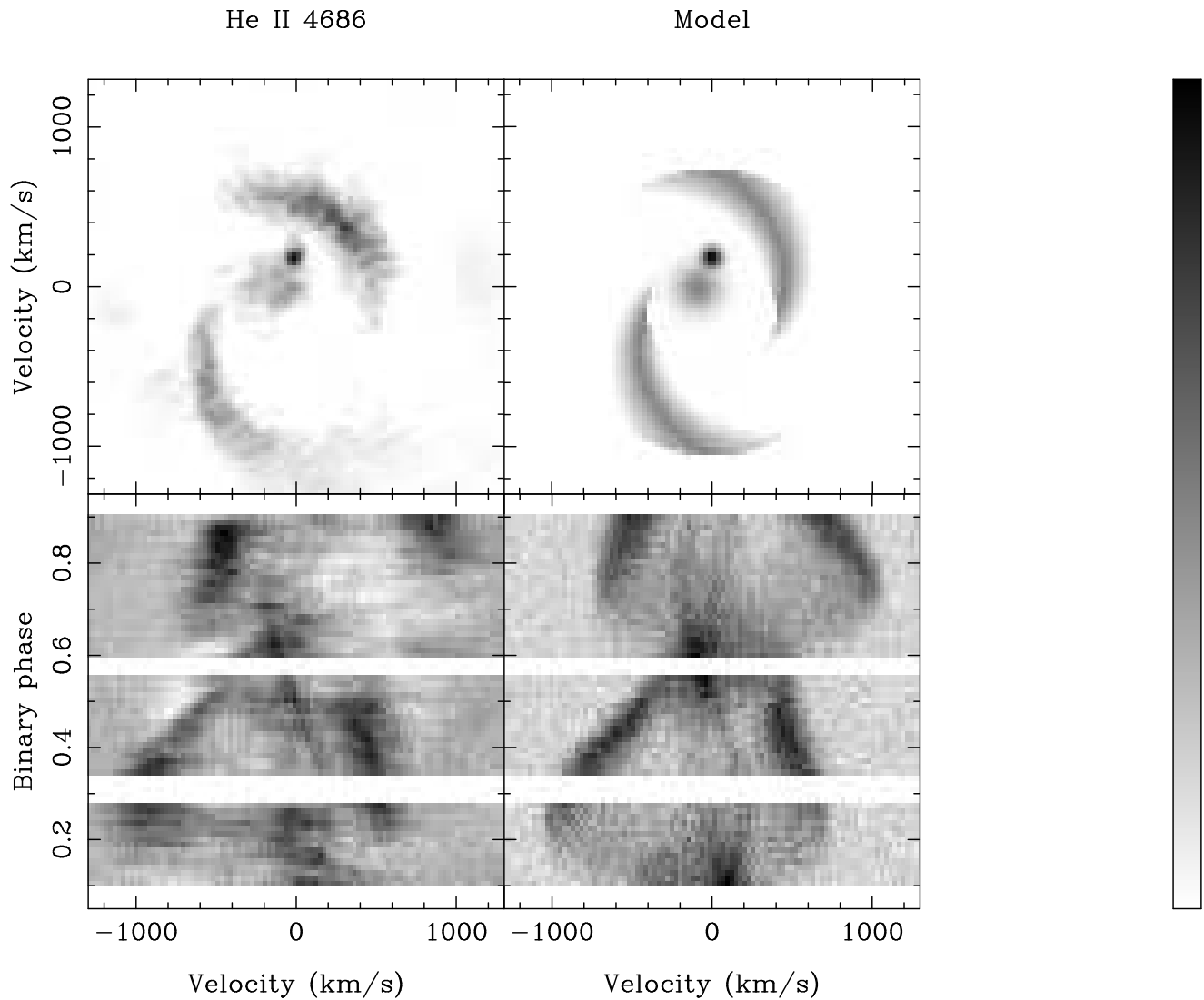}
\caption[]{A model using spiral arms (top-right) is shown
for comparison to the observed non-axisymmetric image (top-left).
The computed data resulting from the model projected at each binary
phase (bottom-right) are also displayed 
for comparison with the non-axisymmetric part of the data (bottom-left). 
The signature of spiral shocks is demonstrated with clarity
on the trailed spectra 
(a decreasing double-peak separation with an intensity jump at phase 0.6).
The greyscale bar corresponds to a linear intensity scale of 0-100 for the
images and of 0-25 mJy for the data.}
\end{figure*}

The trailed spectra   of the strongest  emission line,  He{\small~II},
show very  complex  structure  (left  panel  in Fig.   2).   The  peak
separation  changes with phase and is  narrower at phase 0.50. The red
peak of the profile is maximum at phases 0.45 and  0.95 where it has a
velocity of 400 and 650 km s$^{-1}$, respectively. Similarly, the blue
peak has a  clear maximum  at  phase 0.85 and  a  velocity of  500  km
s$^{-1}$.   Other weaker components  are  immediately visible; a sharp
component moves from red to blue at phase 0.5  (red star emission) and
a low-velocity component is present at phases around  0.2 and 0.9.  We
reconstruct the Doppler image of  the He{\small~II} 4686 emission-line
distribution using the  trailed  spectra (Marsh and Horne  1988). This
imaging technique has been  particularly  successful in resolving  the
location of   emission components  such   as the  red   star (IP  Peg;
Harlaftis et al. 1994), the gas  stream (OY Car in outburst; Harlaftis
and Marsh  1996), the bright spot  (GS2000+25; Harlaftis et al.  1996)
and  spiral waves in  the outer accretion disc  (from H$\alpha$ and 
He{\small~I} lines of IP Peg during rise to
outburst;  Steeghs,  Harlaftis and    Horne 1997). 

The He{\small~II}  Doppler  image of  IP  Peg during outburst  (centre
panel  of  Fig.  2)  directly displays the   locations of  the various
emission  components; the inner side of  the  red star, a low-velocity
component and the  dominant accretion disc   with  extended  spiral
arms.  For comparison with the observed data, the computed data from the
image are also shown (right panel).
 The     low-velocity   component  is      centred  at $(V_{x},
V{y})=(-100\pm50,-20\pm70)$ km  s$^{-1}$   and has a  FWHM   of 270 km
s$^{-1}$ and although is seen in other dwarf novae during outburst its
origin is not understood (Steeghs et al.  1996).  In Fig. 3 we display
the  non-axisymmetric data  and Doppler  image for   comparison with a
model.   The non-axisymmetric Doppler   image  (top-left) is built  by
subtracting the  median at each  radius  with the  white dwarf  at its
centre.   A simple  model of  spiral   arms (plus a   red  star and  a
low-velocity component; top-right)  is built by using  a  power law of
$V^{-1}$ for the intensity, velocities between  400 to 800 km s$^{-1}$
and  azimuthal extent of 0.25  binary cycles.   The computed data from
the model (bottom-right;  with the  same  noise content as  the  data)
reconstruct  successfully all the velocity  features observed, such as
the decreasing double-peak separation with an  intensity jump at phase
0.6.  The Bowen blend, He{\small~I}  4471 and Mg{\small~II} 4481 lines
produce similar Doppler images.

\begin{figure}
%\vspace{15cm}
%\psdraft
\psfig{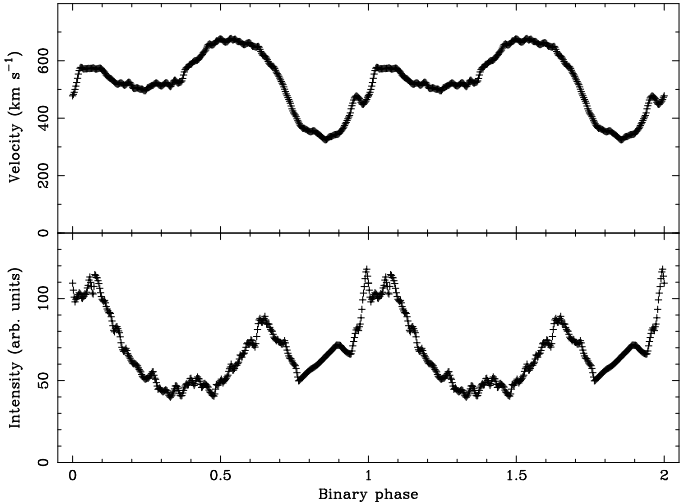}
\caption[]{The variation of maximum intensity and velocity at maximum
intensity with binary phase, as
extracted from the He{\small~II} Doppler image (Fig. 2)
(see also text).}
\end{figure}

\section{Discussion}

The  high-ionization He{\small~II}   4686  line is  most  suitable  to
clarify the shock nature of the resolved spiral structure, since it is
sensitive to higher temperatures and is not saturated as H$\alpha$ may
be.   We quantify the properties  of  the spiral  structure by  taking
radial  slices starting from the white  dwarf at  (0,-147) km s$^{-1}$
for each azimuth and then search for the velocity  for which the image
intensity reaches its  maximum value. The result  is then to trace the
centre of the spiral arms with azimuth, maximum intensity (bottom) and
velocity at that image value, relative to  the white dwarf (top).  The
two peaks  in both velocity   and intensity  correspond to the  spiral
arms.  In particular, the  jump in intensity (a   factor of more  than
two)  and velocity ($\sim$200--300  km s$^{-1}$) is the characteristic
signature for  a  shock.  The  shocks   show an  azimuthal  extent  of
$\sim$90 degrees (phases 0.50--0.75 and 0.95--1.25) and a very shallow
power-law index    in  emissivity  ($\sim~V^{-1}$   or $\sim~R^{-2.5}$
assuming a Keplerian disc).  The spiral shocks contribute about 15 per
cent of  the total disc emission  and their opening angle is $\sim~30$
degrees by assuming Keplerian flow (which places  a firm upper limit).
The Mach numbers  at the shocks range between  28--50 for $T=20,000$ K
and velocities between  400--700  km s$^{-1}$.  There  are differences
between the two  arms: the arm centred at  phase 0.65 (bottom-left arm
on Doppler image) is weaker in intensity (by 40 per cent) than the arm
centred  at phase  0.05.  Also,   the  velocity range  is 400--700  km
s$^{-1}$ compared to  400--550 km s$^{-1}$  and the radial width is 40
per  cent shorter compared to the  FWHM=240 km s$^{-1}$  of the arm at
phase 0.05.   Note that in the  above we have discussed the velocities
related to the  maximum  intensities of  the image.  The Doppler image
shows that there is  also some emission  at velocities  up to 1000  km
s$^{-1}$ (or 850 km s$^{-1}$ relative to the white dwarf).

Spiral arms have now been observed in  3 different outbursts of IP Peg
(this work; 1993 August outburst in Steeghs, Harlaftis and Horne 1997;
1987 July outburst  in Marsh \& Horne  1990 where, in retrospect, only
limited  evidence  for spiral structure   can be discerned  due to the
incomplete binary coverage).   The trailed spectra obtained by Hessman
(1989) immediately after the end of an outburst  of IP Peg do not show
any structure  other than the red  star `S'-wave component.  We tested
this by adjusting our He{\small~II} spectra to the instrument and time
resolution of the Hessman spectra (195 km s$^{-1}$  and 12 phase bins,
respectively). The signal-to-noise is similar  since the red star line
component is visible. The spiral  shocks can still  be resolved in the
simulated  trailed  spectra suggesting   that  the shocks  either  are
tightly wound up  or may have  dissipated  by the time the  system has
reached  quiescence after an  outburst.   Spiral structure in  systems
other than IP Peg can also be found, such as in SS Cyg during outburst
(see Helium lines in Figures 7, 9 and 10; Steeghs et al.  1996) and in
the H$\alpha$ trailed spectra of LB1800 (Still  et al.  1998, Steeghs
et al.   1998, in  preparation).   No  evidence  of spiral  shocks  is
observed  in  short-period SU  UMa-type stars, such   as OY Car during
outburst (Harlaftis and Marsh 1996), SU UMa and  YZ Cnc in outburst or
quiescence  (Harlaftis 1991).  Therefore, it may  be possible that the
spiral shocks  only develop in systems  above the period gap where the
disc,  the companion star and   the  mass  transfer rate are   larger.
Doppler  maps of  X-ray  binaries do not   show any evidence of spiral
shocks  (outburst of GRO~J0422+32, Casares et  al. 1995; quiescent map
of A0620-00, Marsh et  al.  1994; quiescent  map of GS~2000, Harlaftis
et al.  1996).  Only  H$\alpha$  spectra of X~1822-371, the  prototype
system  for  vertical disc  structure,  may  contain  a  hint  of some
structure similar to spiral shocks (Harlaftis et al. 1997).

SPH simulations of  the IP Peg accretion  disc do show the development
of   transient spiral shocks (Armitage and    Murray 1998) which agree
quite well  with the observed ones  in  velocity and azimuthal extent.
Stehle (1998) uses hydrodynamical simulations  of a hot accretion disc
($T_{eff}=10,000-50,000$  K, $q=0.3$) which  confirm the deployment of
stable spiral shocks in the  disc over many  orbital cycles (see  also
Steeghs  and Stehle 1999).  Fitting  of the He{\small~II} spiral shock
data  (Fig.  4) with model  data extracted from simulated discs should
eventually constrain the  physical conditions of  the disc such as the
local Mach number and effective  temperatures. We are aiming to obtain
similar coverage at the end of an outburst (to complete our coverage),
so that we have a full picture of the dynamical evolution of the tidal
shocks.  In particular, the effect  of a decreasing $\alpha$ parameter
(outburst decline) on the velocity  range, opening angle and azimuthal
extent of the spiral shocks.

\section*{Acknowledgments}

The   data reduction  and   analysis was partly   carried   out at the
St. Andrews STARLINK node.  Use of  software developed by T.  Marsh is
acknowledged.   ETH    was   supported  by    the   TMR  contract   RT
ERBFMBICT960971 of  the European  Union.  ETH  and DS   were partially
supported by  a  joint research programme   between the  University of
Athens,    the British  Council at  Athens    and  the University   of
St. Andrews.  DS  acknowledges support by a  University of St. Andrews
research studentship. In this research,  we have used, and acknowledge
with  thanks,  data from the  AAVSO   International Database, based on
observations   submitted  to the   AAVSO  by  variable star  observers
worldwide.

\label{lastpage}

\end{document}